\documentclass[a4paper,11pt]{article}
\usepackage{pos}
\usepackage{siunitx}
\usepackage{subfigure}

\PassOptionsToPackage{hyphens}{url}\usepackage{hyperref}

\title{Mu3e Integration Run 2021}

\author*{Marius Köppel on behalf of the Mu3e collaboration}
\affiliation{
    Institute for Nuclear Physics and PRISMA+ Cluster of Excellence,\\
    Johann-Joachim-Becher-Weg 45, Mainz, Germany
}

\emailAdd{mkoeppel@uni-mainz.de}

\abstract{The Mu3e experiment at the Paul Scherrer Institute searches for the charged lepton flavor violating decay $\mu^+ \rightarrow e^+ e^+ e^-$.
The experiment aims for an ultimate sensitivity of one in $10^{16}$ $\mu$ decays. 
The first phase of the experiment, currently under construction, will reach a branching ratio sensitivity of $2.5\times10^{-15}$ by observing $10^{8}$ $\mu$ decays per second over a year of data taking. 
The highly granular detector based on thin high-voltage monolithic active pixel sensors (HV-MAPS) and scintillating timing detectors will produce about 100 GB/s of data at these particle rates.
The Field Programmable Gate Array (FPGA) based Mu3e Data Acquisition System will read out this data from the detector and reduce the event rate to 100 MB/s by selecting interesting events using a filter farm of graphics processing units.
The paper presents the status of the data acquisition system (DAQ) and first results from the 2021 integration run, which for the first time operated a slice of the Mu3e detector at the $\pi$E5 muon beam line at the Paul Scherrer Institute (PSI).}

\FullConference{%
  *** The 22nd International Workshop on Neutrinos from Accelerators (NuFact2021) ***\\
  *** 6–11 Sep 2021 ***\\
  *** Cagliari, Italy ***}

\begin{document}
\maketitle

\section{The Mu3e Experiment}
Observing lepton flavor violation in charged leptons would be a clear hint for new physics.
The Mu3e Experiment~\cite{ARNDT2021165679}, which is currently under construction at PSI in Switzerland, could improve the current limit of $\mathcal B (\mu^+\rightarrow e^+e^+e^-) < \SI{1.0 e-12}{}$, set by the SINDRUM experiment~\cite{sindrum}, by three orders of magnitude.
To this end, a muon rate of $\SI{e8}{\mu / s}$ and data taking of more than one year is needed.

Mu3e uses a~\SI{29.8}{MeV/c} muon beam produced by the decays of pions on the surface of a graphite target hit by the~\SI{590}{MeV} proton beam at PSI.
The muons are guided into a~\SI{1}{T} solenoid magnet and stopped on a hollow double cone target made out of Mylar foil.
The trajectories of the decay products of the stopped muons will be measured using a silicon pixel tracking detector.
The strong magnetic field forces the trajectories of particles to curl back into the detector, which increases the overall lever arm of the momentum measurement.

\begin{figure}[ht]
\centering
\includegraphics[scale=0.35]{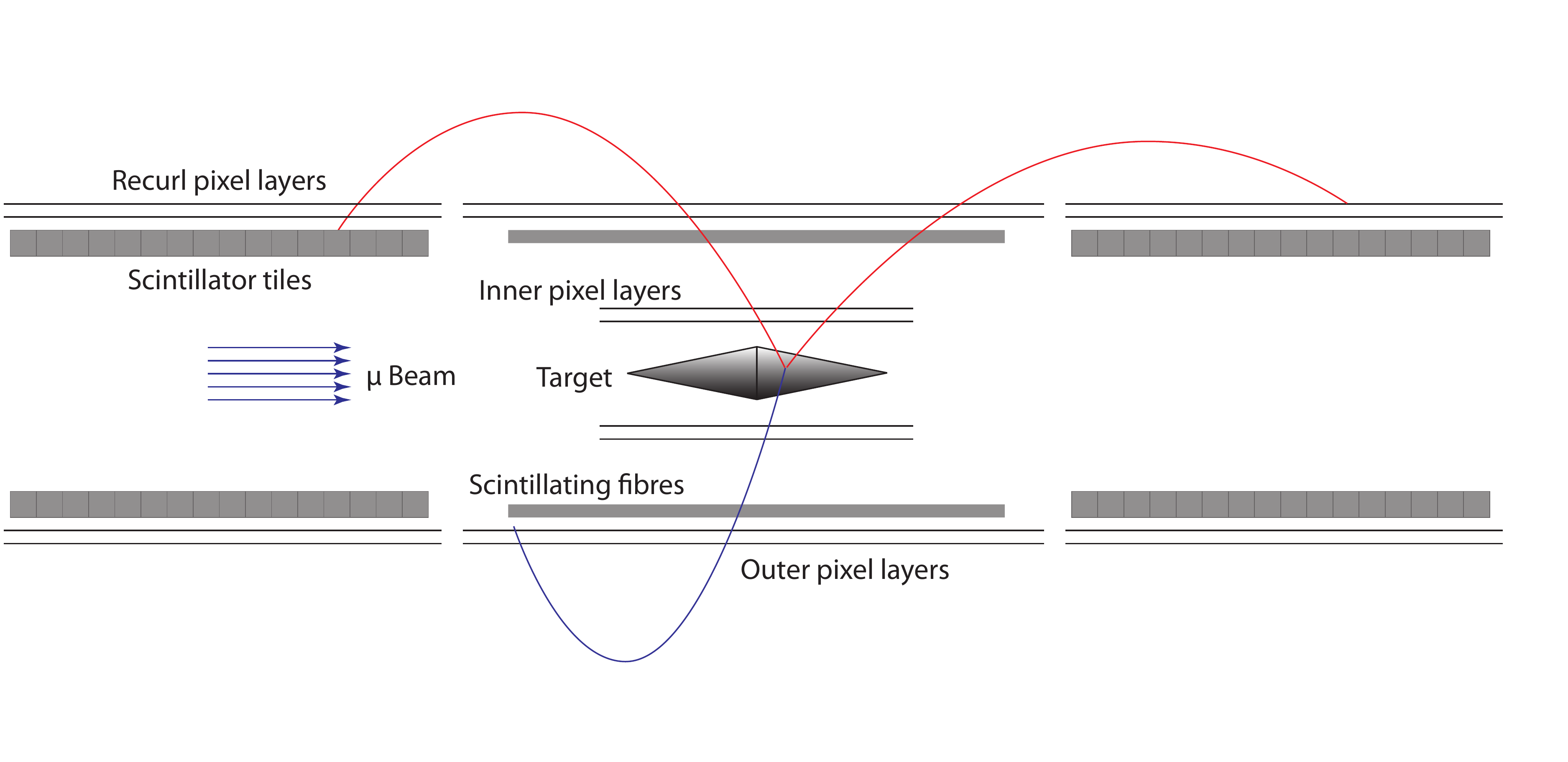}
\caption{Sketch of the Mu3e detector concept.}
\label{fig:detector}
\end{figure}
In Figure~\ref{fig:detector}, a sketch of the Mu3e detector is shown.
The detector is built in three parts, one inner central station and two recurl stations, one up- and one downstream.

Multiple scattering is the biggest contribution to the momentum resolution.
Therefore, a new type of thin high-voltage monolithic active pixel sensors, called MuPix, is used, which can be thined to~\SI{50}{\mu m}~\cite{mupix}.
For precise timing measurements a scintillating fibre detector, placed in the central station of the detector, and scintillating tile detectors, located in the two recurl stations, are used.
These detectors are read out via silicon photomultipliers and a custom designed application specific integrated circuit, called MuTRiG~\cite{mutrig}.

\section{The Mu3e Integration Run 2021}\label{sec:intRun}

\begin{figure}[ht]
\centering
\includegraphics[scale=0.35]{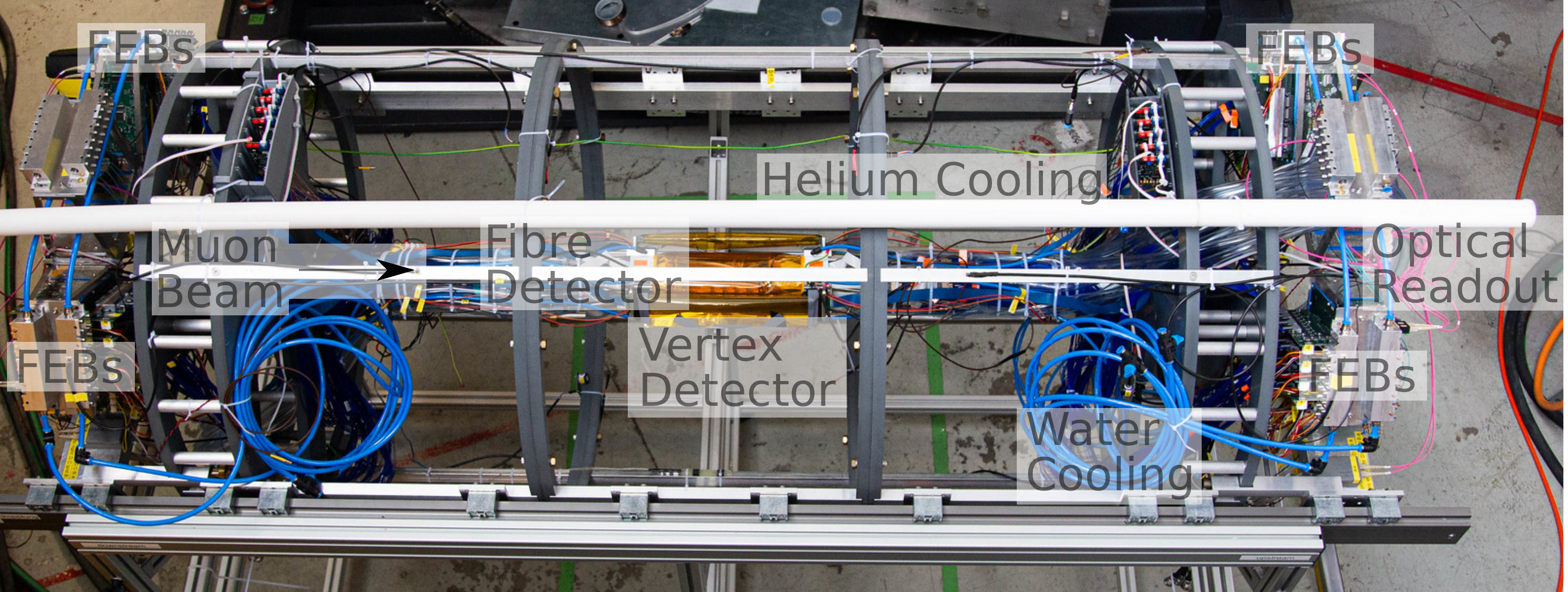}
\caption{Picture of the Mu3e Integration Run 2021 detector prototypes.}
\label{fig:intRun}
\end{figure}
During the Mu3e Integration Run 2021, detector prototypes were tested at the $\pi$E5 muon beam line at PSI.
In Figure~\ref{fig:intRun}, the Integration Run setup is shown.
The intention of the run was to operate the detectors under conditions as close as possible to the final experiment.
This includes cooling of the MuPix detector with gaseous helium, operating the whole setup inside the~\SI{1}{T} solenoid magnet and with a muon beam.
As indicated in Figure~\ref{fig:intRun}, the muon beam is entering the detector from the left.
The vertex detector and the fibre detector are placed in the middle of the cylindrical support frame surrounding the beampipe and the target.
The support frame is holding the different supply chains for cooling, electrical power and the optical readout cables.
The custom developed readout boards are annotated with Front-End Boards (FEBs).
Detector-wise the setup contained the inner vertex detector prototype and two ribbons of the scintillating fibre detector.
Beside the hollow double cone Mu3e target, a second target was inserted to study muon spin rotation ($\mu$SR) measurements.
The second target was made out of two discs.
The first one holds a scintillator used as an entrance counter.
On the second disc the actual target, which was a silver probe inside a permanent magnet, is mounted.
The $\mu$SR measurements were performed with no external magnetic field.

\subsection{The Integration Run Data Acquisition System}
\begin{figure}[ht]
\centering
\includegraphics[scale=0.29]{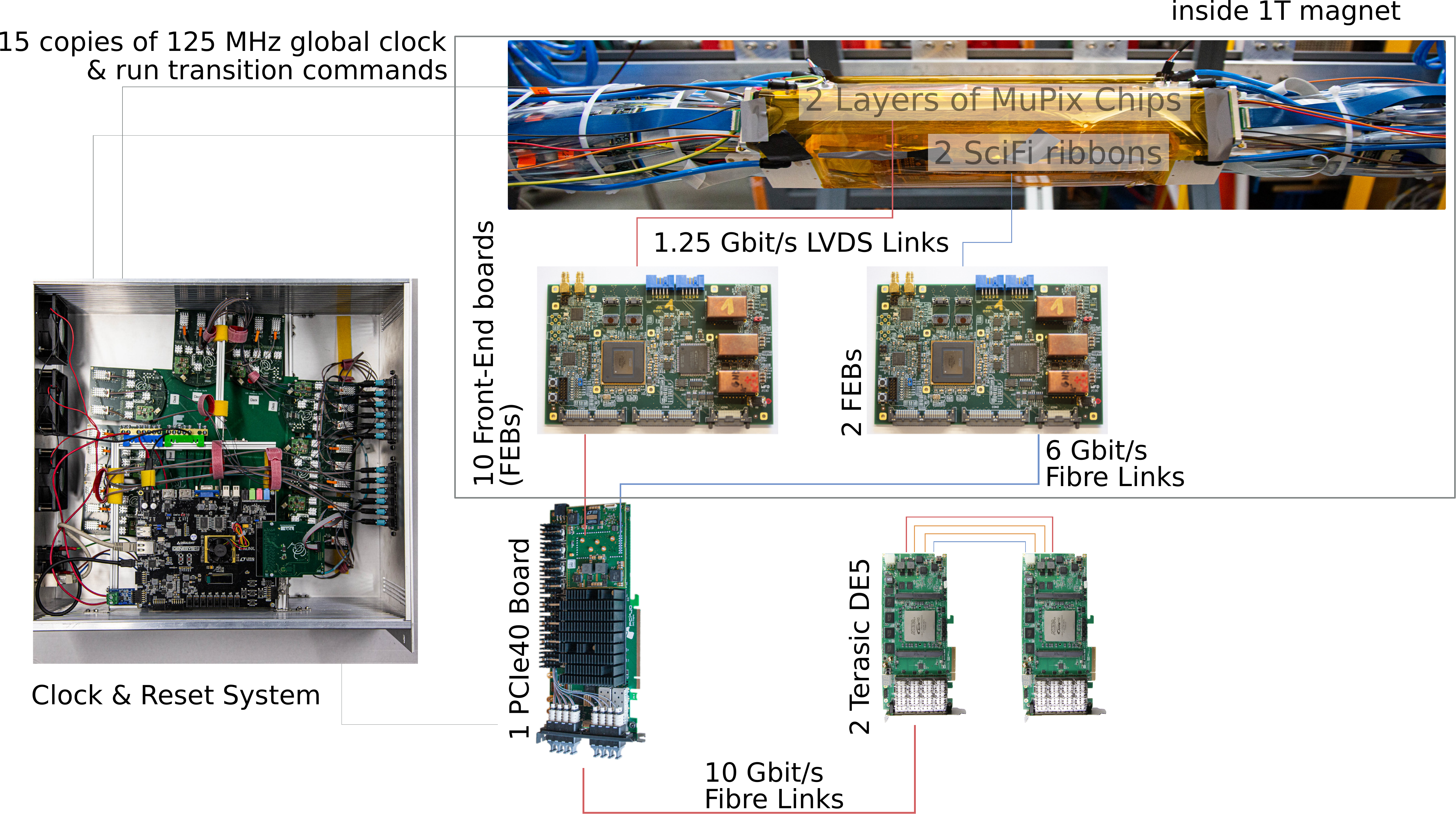}
\caption{Sketch of the Mu3e Integration Run 2021 data acquisition system.}
\label{fig:intRunDAQ}
\end{figure}
In Figure~\ref{fig:intRunDAQ}, a sketch of the DAQ used in the Integration Run is shown.
The DAQ is built up of 3 layers of FPGA boards.
For reading out the different detectors the custom developed FEB is used.
This board is located inside the magnetic field and is connected via electrical~\SI{1.25}{Gbit/s} LVDS links to the different sub-detectors.
The task of the board is to readout and configure the different detectors and sort the received hits in time.
The second layer of the DAQ contains the PCIe40 Board~\cite{pcie40}, which was developed for the ALICE and LHCb upgrades.
These boards are sitting outside the magnet and are connected via~\SI{6}{Gbit/s} optical links to the FEBs.
They are used to combine and time-sort the different data streams and to send the detector configuration via optical connections to the FEBs.
The last layer of the system contains the commercial Terasic DE5a-Net-DDR4 boards~\cite{de5net} sitting in a PC equipped with a graphics processing unit (GPU).
The onboard DDR4-RAM is used to buffer the data while the hits of the two layers of the central detector are used to perform an online event selection using the GPU.
In the final system twelve of these so called farm PCs are daisy-chained to process the total amount of~\SI{100}{GB/s} of data.
During the Integration Run the GPU selection was not used.
To synchronize the different detectors a dedicated clock and reset system is used.
This contains another FPGA board which generates a~\SI{125}{MHz} clock with a synchronized reset for run starts and stops.
A full description of the final DAQ of the Mu3e detector is given in~\cite{daq}.

\subsection{First Results of the Mu3e Integration Run 2021}
To verify that all subsystems and the DAQ are working, various correlations are studied.
Since the scintillating fibre detector was not fully working during the data taking, only the results for the pixel vertex detector are shown.

\begin{figure*}[!ht]
\centering
\subfigure[Column correlations]{\label{fig:space_cor_mu3e}\includegraphics[width=0.36\textwidth]{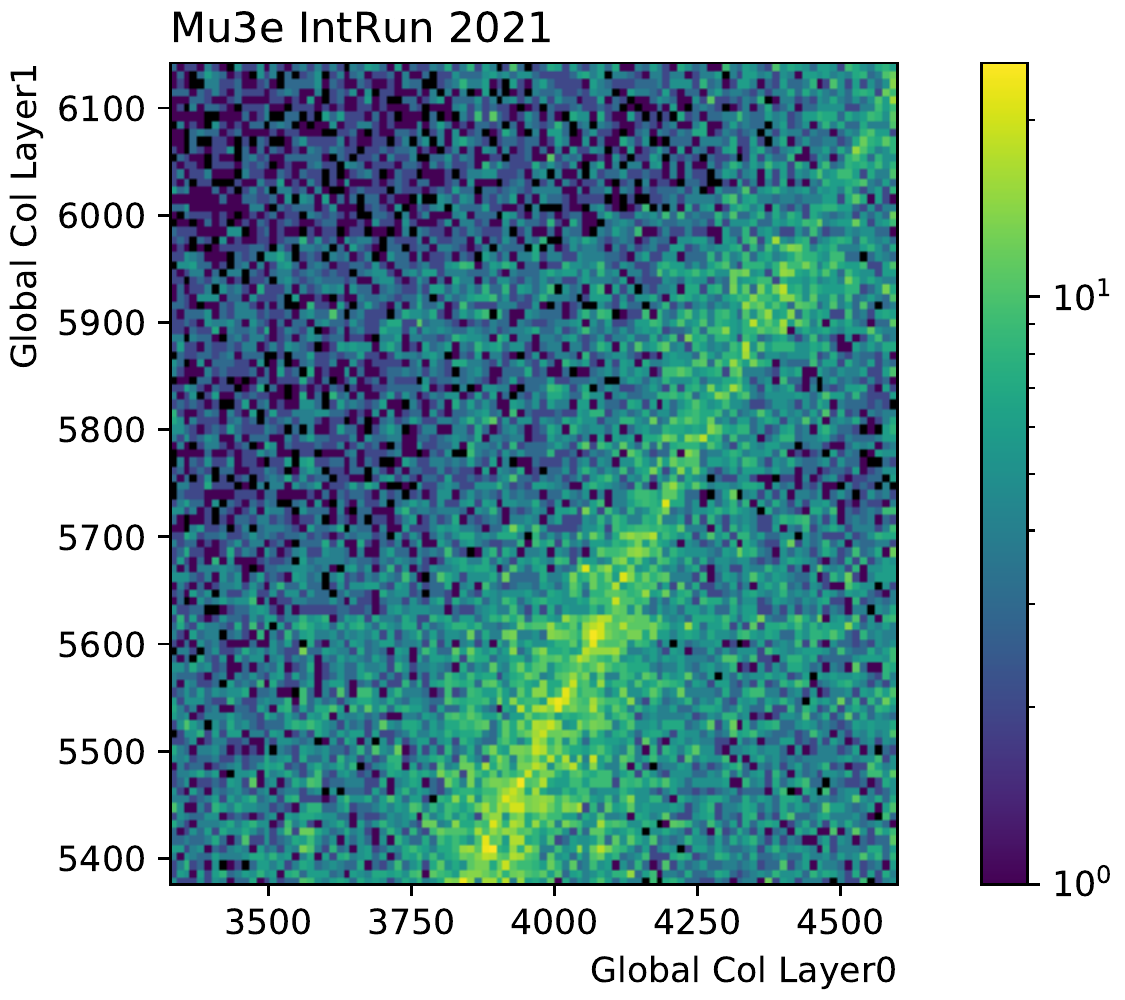}}
\subfigure[Time difference]{\label{fig:time_diff_mu3e}\includegraphics[width=0.36\textwidth]{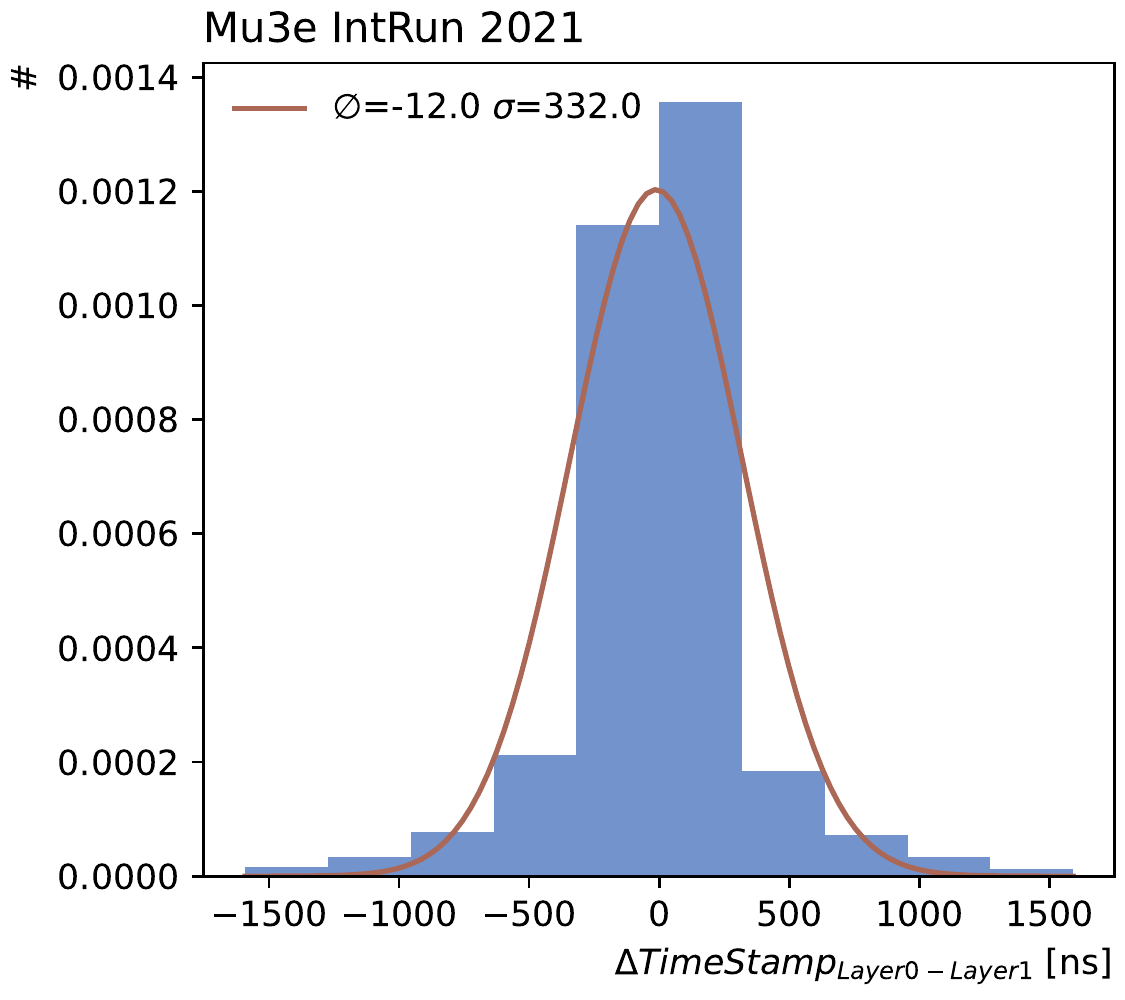}}
\caption{Column to column correlations of the two layers of the vertex detector prototype are shown in Figure~\ref{fig:space_cor_mu3e}. The positions of the different pixel hits are indicated with global column positions of the individual pixels on each layer of the vertex detector. Time difference of two pixel hits in the two layers of the vertex detectors is displayed in Figure~\ref{fig:time_diff_mu3e}.}
\label{fig:intRunResults}
\end{figure*}
In Figure~\ref{fig:intRunResults} space and time correlations of the pixel chips are displayed.
During the run only parts of vertex detector were running.
Figure~\ref{fig:space_cor_mu3e} shows a part of the vertex detector where a clear correlation between the two layers is visible.
The data was taken with magnetic field turned on.

Besides the space correlation in Figure~\ref{fig:time_diff_mu3e} the time difference of two consecutive pixel hits, in the two layers of the vertex detector, is shown.
Each pixel hit sends beside its space information a time information.
The time difference shows a sigma of \SI{332}{ns}.
Taking into account that the detector was not fully calibrated, the measured resolution is in line with the results observated in various MuPix testbeams~\cite{mupix10}.

\subsection{First Results of the $\mu$SR measurement}
For the $\mu$SR measurement the target was replaced and the same correlations studies are performed.

\begin{figure}[ht]
\centering
\includegraphics[width=0.36\textwidth]{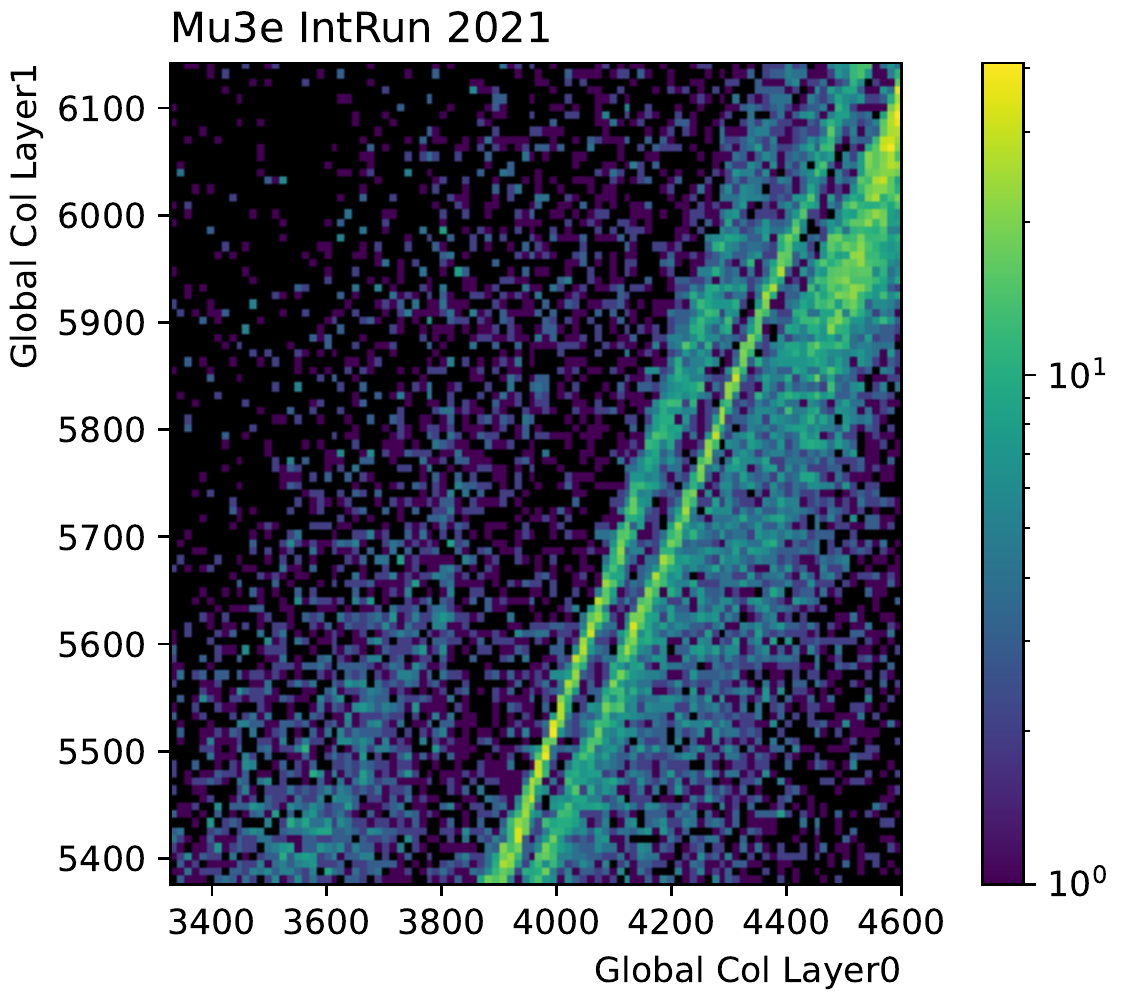}
\caption{Column to column correlations of the two layers of the vertex detector prototype using the $\mu$SR target are shown. The positions of the different pixel hits are indicated with global column positions of the individual pixels on each layer of the vertex detector.}
\label{fig:space_cor_musr}
\end{figure}
The space correlations of the inner vertex detector are shown in Figure~\ref{fig:space_cor_musr}.
In contrast to the Mu3e target the space correlation of the two layers of the vertex detector show two lines.
As described in~\ref{sec:intRun} the target was built out of two discs which are a few centimeters apart from each other.
The two lines in the space correlation are corresponding two the different discs of the target.
More details about the whole measurement can be found in~\cite{thomas}.

\section{Summary and Outlook}
With the Mu3e integration run 2021 a milestone on the roadmap of commissioning the full detector was achieved.
Various systems showed that they work as intended and the vertex detector prototype showed reasonable results while running under the final environmental conditions of the experiment.
In 2022 further commissioning runs are planed to fully integrate the other sub-detectors.
Further integration of the readout system will be carried out to finalize the data acquisition system leading to first production data taking in 2024.

%
%

\end{document}